\documentclass[fleqn,a4paper,12pt]{article}
\usepackage{amssymb}
\usepackage{makeidx}
\usepackage{amsmath}
\usepackage{graphicx}
\usepackage{lscape}
\usepackage{a4}
\usepackage{epsfig}
\begin{document}

\thispagestyle{empty}
\newcommand{\al}{\alpha}
\newcommand{\bt}{\beta}
\newcommand{\s}{\sigma}
\newcommand{\lbd}{\lambda}
\newcommand{\vp}{\varphi}
\newcommand{\va}{\varepsilon}
\newcommand{\gm}{\gamma}
\newcommand{\G}{\Gamma}
\newcommand{\p}{\partial}
\newcommand{\om}{\omega}
\newcommand{\be} {\begin{equation}}
\newcommand{\lo}{\left(}
\newcommand{\ro} {\right)}
\newcommand{\ee} {\end{equation}}
\newcommand{\ba} {\begin{array}}
\newcommand{\ea} {\end{array}}
\newcommand{\ds}{\displaystyle}

\begin{center}
{\Large Relativistic Coulomb problem for particles with arbitrary
half-integer spin}
\end{center}
\vspace{2mm}
\begin{center}
J. Niederle\\
Institute of Physics of Acad. Sci. of the Czech Republic, Na
Slovance 2, 18840 Prague,
Czech Republic,\\
  A. G. Nikitin\\

Institute of Mathematics of Nat.Acad. Sci of Ukraine, 4
Tereshchenkivska str. 01601 Kyiv, Ukraine
\end{center}
\vspace{2mm} {\bf Abstract} Using relativistic  tensor-bispinorial
equations proposed in \cite{NN} we solve the Kepler problem for a
charged particle with arbitrary half-integer spin interacting with
the Coulomb potential.

\section{ Introduction}

Exactly solvable problems in quantum mechanics are quite important
and illustrative but rather rare (see, e.g., \cite{bagr}).  They
can be described fully and in a straightforward way free of
various complications caused  by the perturbation method. The very
existence of exact solutions of these problems is usually
connected with their non-trivial symmetries which are mostly of
particular interest by themselves In addition exact solutions form
complete sets of functions which can be used to find
  solutions of other
problems.

One of the triumphs of the Dirac theory for the electron is that
it solves the problem of electron motion in the Coulomb potential.
The exact Sommerfeld formula for the related energy levels is one
of the cornerstones of relativistic quantum theory. However
extension of these results to the case of charge particles with
higher spin appeared to be very complicated since even in the case
of spin $s=1$ the corresponding relativistic wave equation
(namely, the Kemmer-Duffin equation) is not free of inconsistences
and predicts  the orbital particle will fall down on the
attracting center \cite{tamm}, \cite{schwin}. The other well known
problems of RWE for particles with higher spins are: violation of
causality \cite{wightman},  ill-defined interaction with a
constant and homogeneous external magnetic field \cite{Tsai}, and
so on (for details see \cite{NN}, \cite{sud}).

In recent paper \cite{NN} relativistic wave equations for
particles of arbitrary half-integer spin were proposed which are
free of inconsistencies typical for other wave equations for
higher spins. They are causal and allow the correct value of the
gyromagnetic ratio $g=2$. In addition these equations have well
defined quasi-relativistic limits which admit a good physical
interpretation and describe the Pauli, spin-orbit and Darwin
couplings.

In the present paper we apply the tensor-spinorial equations
\cite{NN} to solve the Coulomb problem for particles with {\it
arbitrary} half-integer spin.

\section{Relativistic wave equations for a particle with arbitrary half-integer spin}
 First, we briefly review the RWE  for  free particles with
arbitrary half-integer spin $s$ and mass $m$ described in
\cite{NN}. The corresponding  wave function $\psi^{[\mu_1\nu_1]
[\mu_2\nu_2]\cdots[\mu_n\nu_n]}$ is an irreducible tensor with
respect to the complete Poincar\'e group of rank $2n=2s-1$
antisymmetric w.r.t. permutations of indices in the square
brackets and symmetric w.r.t. permutations of pairs of indices
$[\mu_i,\nu_i]\rightleftarrows[\mu_j,\nu_j],\ i,j=1,2,\cdots,n$.
The irreducibility requirement means also that convolutions w.r.t.
any pair of indices and cyclic permutations of any triplet of
indices of the wave function reduce it to zero. In addition,
components of $\psi^{[\mu_1\nu_1] [\mu_2\nu_2]\cdots[\mu_n\nu_n]}$
are bispinors of rank 1. This means that the wave function has an
additional spinorial index $\al$ (which we omit) running from 1 to
4.

The equation of motion has the following form \cite{NN}
\begin{equation}\label{2.2}\ba{l}
    \left(\gamma_\lambda p^\lambda-m\right)\psi^{[\mu_1\nu_1]
[\mu_2\nu_2]\cdots[\mu_n\nu_n]}\\ \ \\-\frac{1}{4s}\Sigma_{\cal
P}\left(\gamma^{\mu_1}\gamma^{\nu_1} -\gamma^{\nu_1}\gamma^{\mu_1}
\right)p_\lambda\gamma_\s\psi^{[\lambda\s][\mu_2\nu_2]
[\mu_3\nu_3]\cdots[\mu_n\nu_n]}=0.\ea
\end{equation}
Here $\gamma_\nu$ are Dirac matrices, $p^\mu=i\frac{\p}{\p x_\mu}$
and the symbol $\Sigma_{\cal P}$ denotes the sum over all possible
permutations of subindices $(2,\cdots,n)$ with 1.

In addition, the wave function $\psi^{[\mu_1\nu_1]
[\mu_2\nu_2]\cdots[\mu_n\nu_n]}$ has to satisfy the static
constraint \cite{NN}
\begin{equation}\label{2.1}
    \gamma_\mu\gamma_\nu\psi^{[\mu\nu]
[\mu_2\nu_2]\cdots[\mu_n\nu_n]}=0
\end{equation}
which is necessary to reduce the number of independent components
of the tensor-spinor from $16s$ to $4(2s+1)$. In order to obtain a
theoretically recognized $2(2s+1)$-component wave function we
impose on $\psi^{[\mu_1\nu_1] [\mu_2\nu_2]\cdots[\mu_n\nu_n]}$
additionally  either Majorana condition or a parity-violating
constraint $(1+i\gm_5) \psi^{[\mu_1\nu_1]
[\mu_2\nu_2]\cdots[\mu_n\nu_n]}=0$ (or $(1-i\gm_5)
\psi^{[\mu_1\nu_1] [\mu_2\nu_2]\cdots[\mu_n\nu_n]}=0$).

Equations (\ref{2.2}), (\ref{2.1}) are manifestly invariant with
respect to the complete Poincar\'e group and admit the Lagrangian
formulation. For the case of a charged particle interacting with
an external electromagnetic field this equation is generalized to
the following form \cite{NN}:
\begin{equation}\label{2.3}\ba{l}
    \left(\gamma_\lambda \pi^\lambda-m\right)\psi^{[\mu_1\nu_1]
[\mu_2\nu_2]\cdots[\mu_n\nu_n]}\\\\-\frac{1}{4s}\Sigma_{\cal
P}\left(\gamma^{\mu_1}\gamma^{\nu_1} -\gamma^{\nu_1}\gamma^{\mu_1}
\right)\pi_\lambda\gamma_\s\psi^{[\lambda\s][\mu_2\nu_2]
[\mu_3\nu_3]\cdots[\mu_n\nu_n]}\\ \ \\+\frac{2iek}{sm}\Sigma_{\cal
P}\lo\frac14\gm_\al\gm_\s F^{\al\s}\psi_+^{[\mu_1\nu_1]
[\mu_2\nu_2]\cdots[\mu_n\nu_n]}\right.\\ \ \\\left.+
{F_\al}^{\mu_1}\psi_+^{[\nu_1\al][\nu_2\mu_2]\cdots[\nu_{n}\mu_{n}]}
-{F_\al}^{\nu_1}\psi_+^{[\mu_1\al][\nu_2\mu_2]\cdots[\nu_{n}\mu_{n}]}\ro=0.\ea
\end{equation}
Here $\pi_\lambda=p_\lambda-eA_\lambda,\ A_\lambda$ and
$F^{\mu\nu}$ are vector-potential and tensor of the external
electromagnetic field, $k$ is an arbitrary parameter and
\[\psi_\pm^{[\mu_1\nu_1]
[\mu_2\nu_2]\cdots[\mu_n\nu_n]}=\psi^{[\mu_1\nu_1]
[\mu_2\nu_2]\cdots[\mu_n\nu_n]}\pm \frac{1}{2n}\gm_5\Sigma_{\cal
P}{\varepsilon^{\mu_1\nu_1}}_{\lambda\s}\psi^{[\lambda\s][\mu_2\nu_2]
[\mu_3\nu_3]\cdots[\mu_{n}\nu_{n}]},\] so that
\[ \psi^{[\mu_1\nu_1]
[\mu_2\nu_2]\cdots[\mu_n\nu_n]}=\frac12\left(\psi_-^{[\mu_1\nu_1]
[\mu_2\nu_2]\cdots[\mu_n\nu_n]}+\psi_+^{[\mu_1\nu_1]
[\mu_2\nu_2]\cdots[\mu_n\nu_n]}\right).\]

Equation (\ref{2.3}) contains both the minimal and anomalous
interaction of a particle with an external field \cite{NN}.
Moreover, putting $k=2s-1$ for the anomalous coupling constant $k$
we get the "natural" value $g=2$ for the gyromagnetic ratio
\cite{NN}.

Solving equation (\ref{2.3}) for $\psi_-^{[\mu_1\nu_1]
[\mu_2\nu_2]\cdots[\mu_n\nu_n]}$ we obtain
 \be\label{2.5}  \psi_-^{[\mu_1\nu_1]
[\mu_2\nu_2]\cdots[\mu_n\nu_n]}=\frac{1}{m}
\gamma_\lambda\pi^\lambda\psi_+^{[\mu_1\nu_1]
[\mu_2\nu_2]\cdots[\mu_n\nu_n]},
\end{equation}
and then by substituting this expression into (\ref{2.3}) we get
the following second order equation:
\begin{equation}\label{2.4}\ba{l}
\lo\pi_\mu\pi^\mu-m^2-\frac{i(k+1)}{2s}S_{\mu\nu}F^{\mu\nu}\ro\psi_+^{[\mu_1\nu_1]
[\mu_2\nu_2]\cdots[\mu_n\nu_n]}=0.\ea\ee Here $S_{\mu\nu}$ are
generators of the Lorentz group whose action on the antisymmetric
tensor-spinor $\psi_+^{[\mu_1\nu_1]
[\mu_2\nu_2]\cdots[\mu_n\nu_n]}$ is given by the following formula
\begin{equation}\label{2.6}\ba{l}
    S^{\mu\nu}\psi_+^{[\mu_1\nu_1]
[\mu_2\nu_2]\cdots[\mu_n\nu_n]}=\frac{i}{4}[\gm^\mu,\gm^\nu]\psi_+^{[\mu_1\nu_1]
[\mu_2\nu_2]\cdots[\mu_n\nu_n]}\\+i\Sigma_{\cal P}\lo
g^{\mu\mu_1}\psi_+^{[\nu\nu_1][\nu_2\mu_2]\cdots[\nu_n\mu_n]}
-g^{\nu\mu_1}\psi_+^{[\mu\nu_1][\nu_2\mu_2]\cdots[\nu_n\mu_n]}\right.\\\left.
-g^{\mu\nu_1}\psi_+^{[\nu\mu_1][\nu_2\mu_2]\cdots[\nu_n\mu_n]}
+g^{\nu\nu_1}\psi_+^{[\mu\mu_1][\nu_2\mu_2]\cdots[\nu_n\mu_n]}\ro,\ea
\end{equation}
where $g^{\mu\nu}$ is the metric tensor with signature
$(+,-,-,-)$.

 In accordance with its
 definition tensor $\psi_+^{[\mu_1\nu_1]
[\mu_2\nu_2]\cdots[\mu_n\nu_n]}$ transforms according the
representation $D(s-1/2,0)\otimes D(1/2, 0)\oplus
D(0,s-1/2)\otimes D(0,1/2)\equiv D(s,0)\oplus D(s-1,0)\oplus
D(0,s)\oplus D(0,s-1)$ of the Lorentz group. Moreover, condition
(\ref{2.1}) reduces this representation to $D(s,0)\oplus D(0,s)$
whose generators without loss of generality can be expressed as
$$S_{ab}=\varepsilon_{abc}S_c,\  \  S_{0a}=i\hat\va S_a, \ a,b=1,2,3,$$ where
$S_a$ are matrices forming a direct sum of two irreducible
representation $D(s)$ of algebra $so(3)$, $\varepsilon_{abc}$ is
totally antisymmetric unit tensor of rank $3$ and $\hat\va$ is an
involutive matrix distinct from the unit one and commuting with
$S_a$. This matrix can be expressed via the Casimir operators
$C_1=S_{\mu\nu}S^{\mu\nu}$ and $
C_2=\frac14\va_{\mu\mu\lbd\s}S^{\mu\nu}S^{\lbd\s}$ of the Lorentz
group: \be\label{va}\hat\va=C_1C_2^{-1}.\ee

 Thus instead of
(\ref{2.4}) we can study the equivalent equation
\begin{equation}\label{2.62}
    \ba{l}
\lo\pi_\mu\pi^\mu-m^2-\frac{i(k+1)}{2s}\hat\va S_a(iF^{0a}+
\frac12\varepsilon_{0abc}F^{bc})\ro\Psi=0\ea
\end{equation}
where $\Psi$ is a $2(2s+1)$-component spinor belonging to the
space of irreducible representation $D(s,0)\oplus D(0,s)$ of the
Lorentz group, $S_a\ (a=1,2,3)$ are direct sums of two
$(2s+1)\times(2s+1)$ matrices forming the irreducible
representation $D(s)$ of algebra $so(3)$. The components of the
related tensor $\psi_+^{[\mu_1\nu_1]
[\mu_2\nu_2]\cdots[\mu_n\nu_n]}$ can be expressed via components
of spinor $\Psi$ by using Clebsh-Gordon coefficients (see Section
7).

Taking into account commutativity of matrix $\hat\va$ with
matrices $S_a$ equation (\ref{2.6}) can be decoupled to two
subsystems
\begin{equation}\label{2.61}
    \ba{l}
\lo\pi_\mu\pi^\mu-m^2-\frac{i(k+1)}{2s}\va S_a(iF^{0a}+
\frac12\varepsilon_{0abc}F^{bc})\ro\Psi_\va=0,\ \ \va=\pm 1\ea
\end{equation}
where $\Psi_\va$ are eigenvectors of $\hat\va$ corresponding to
the eigenvalues $\va=\pm 1$.

Thus in spite of a rather complicated form of the first order
equations (\ref{2.3}) we received the second order equation
(\ref{2.4}) for $\psi_+^{[\mu_1\nu_1]
[\mu_2\nu_2]\cdots[\mu_n\nu_n]}$ and relation (\ref{2.5}) which
expresses $\psi_-^{[\mu_1\nu_1] [\mu_2\nu_2]\cdots[\mu_n\nu_n]}$
in terms of $\psi_+^{[\mu_1\nu_1]
[\mu_2\nu_2]\cdots[\mu_n\nu_n]}$. Moreover, equation (\ref{2.4})
can be reduced to (\ref{2.61}). This equation has been already
solved for particle interacting with a constant and homogeneous
magnetic field \cite{NikGal}. We shall see that equations
(\ref{2.61}) are very convenient and are easy to handle in the
important case when the external field is generated by a point
charge.

\section{Radial equations for the Coulomb problem}

Consider a charged particle with arbitrary half-integer spin $s$
and electric charge $e$ interacting with an external
electromagnetic field. When this field is generated by a point
charge $Ze$ the related vector-potential has the form
\be\label{3.1}{\bf A}=0, \ \ A_0=\frac{\al}{r},\ee
 where $r=\sqrt{x_1^2+x_2^2+x_3^2}$ and $\al=Ze^2$.

 Now we shall use the reduced version of the equations of motion given by
 expressions (\ref{2.4}) to (\ref{2.61}). Since both equations
 (\ref{2.61}) corresponding to $\va=1$ and to $\va=-1$ lead to the same
 energy spectrum we shall consider only the case $\va=1$ and omit index
 $\va$ of function $\Psi_\va$ in the formulae of this section which follow.

   For the
states with energy $E$ the
 corresponding solutions $\Psi$ can be written as
 \be\label{3.2}\Psi=\exp(-iEx_0)\psi(\bf r),\ee
 where $\psi(\bf r)$ is a $(2s+1)$-component function depending on
 spatial variables and satisfying the following second order
 equation
 \be\label{3.3}\lo E-\frac{\al}{r}\ro^2\psi=\lo m^2-\Delta
 +ik\al\frac{\bf S\cdot \bf r}{r^3}\ro\psi.\ee

 Taking into account the rotational invariance of equation
 (\ref{3.3}) it is convenient to expand its solutions in terms of
 spherical spinors $\Omega^s_{j\ l\ m}$:
 \begin{equation}\label{3.4}
    \psi=\xi_\lambda(r)\Omega^s_{j\ j-\lambda \ m},
\end{equation}
  where $\Omega^s_{j\ l\ m}$ are orthonormalized joint
 eigenvectors of the following four commuting operators: of total angular
 momentum square $J^2$, orbital momentum square $L^2$, spin square $S^2$ and
 of the third component of the total angular momentum $J_3$,
 whose  eigenvalues are
 $j(j+1),\ l(l+1),\ s(s+1) $ and $m$ respectively. Denoting
 $l=j-\lambda$ we receive
 $$\ba{l}
m=-j,-j+1,\cdots,j,\ \ \text{and}\ \ \lambda=-s,-s+1,\cdots
-s+2m_{sj},\ea$$
where $m_{sj}=s$ if $s\leq j$ and $m_{sj}=j$ if $s>j$.

The expressions for spherical spinors via spherical functions are
given in the Appendix.

Thus we expand $\psi(\bf x)$ in accordance with formula
(\ref{3.4})
where $\xi$ are radial functions and summation over $\lambda$ is
imposed which takes the values indicated in the above. We note
that the action of the scalar matrix $\bf S\cdot \bf r$ to the
spinors $\Omega_{j\ j-\lambda \ m}$ is well defined and given by
the formula \cite{MH}
\begin{equation}\label{3.5}
    {\bf S}\cdot {\bf r}\ \Omega_{j\ j-\lambda \ m}=
    rK^{sj}_{\lambda\lambda'}\Omega_{j\ j-\lambda' \ m},
\end{equation}
where $K^{sj}_{\lambda\lambda'}$ are numerical coefficients whose
values are presented in the Appendix.

Substituting (\ref{3.4}) into (\ref{3.3}), using (\ref{3.5}) and
the following representation for the Laplace operator $\Delta$
\begin{equation}\label{3.6}
\Delta=\frac{1}{r^2}\lo\frac{\p}{\p x}\lo r^2\frac{\p}{\p
x}\ro-L^2\ro,
\end{equation}
where $L^2$ is the square of the orbital momentum operator ${\bf
L}={\bf r}\times{\bf p}$, we receive the following equations for
the radial functions
\begin{equation}\label{3.7}
    F\xi_\lambda=\frac{1}{r^2}M_{\lambda\lambda'}\xi_{\lambda'},
\end{equation}
where $F$ is the second order differential operator
\begin{equation}\label{3.8}
   F=
   \frac{d^2}{dr^2}+\frac2r\frac{d}{dr}+(E+\alpha/r)^2-m^2-\frac{j(j+1)}{r^2}
\end{equation}
and $M$ is a matrix whose elements are
\begin{equation}\label{3.9}
    M_{\lambda\lambda'}=\lambda(\lambda-2j-1)\delta_{\lambda\lambda'}+ig\alpha
    K_{\lambda\lambda'}^{sj}.
\end{equation}

Formula (\ref{3.7}) presents the equation for the radial wave
function of a particle with arbitrary half-integer spin
interacting with the Coulomb field.

\section{Energy spectrum}

Matrix $M$ is {\it normal}, i.e., it satisfies the condition
$MM^\dag=M^\dag M$. Thus it is possible to diagonalize it using
some invertible matrix $U$: \be\label{n0}M\to {\tilde M}=UMU^{-1},
\ \ {\tilde
M}_{\lambda\lambda'}=\delta_{\lambda\lambda'}\nu_\lambda\ee (where
$\nu_\lambda$ are eigenvalues of $M$) thus  system (\ref{3.7}) is
reduced to the sequence of decoupled equations
\begin{equation}\label{4.1}
F{\tilde\xi}^\lambda=\frac{1}{r^2}\nu_\lambda{\tilde\xi}^\lambda\
\ \text{(no sum over $\lambda$) }
\end{equation}
where $\tilde\xi^\lambda $ is a $\lambda$ component of vector
$\tilde \xi=U\xi$ .

The other interpretation of $\tilde\xi^\lambda $ is as follows.
one. Let $\xi^{(\lambda)}$ be an eigenvector of matrix $M$
corresponding to the eigenvalue $\nu_\lambda$. Then it can be
represented as
\be\label{an}\xi^{(\lambda)}=\tilde\xi^\lambda\om_\lambda\ \
(\text{no sum over}\ \lambda), \ee where $\om_\lambda$ is the
eigenvector of $M$ which does not depend on $\rho$ and
$\tilde\xi^\lambda$ is a multiplier depending on $\rho$.

Changing the variables $r\to\rho=2\sqrt{m^2-E^2}r,\ \tilde\xi\to
f=\sqrt\frac{\rho}{m^2-E^2}\tilde\xi$ equation (\ref{4.1}) is
transformed to the well known form
\begin{equation}\label{4.2}
    \rho\frac{d^2
    f}{d\rho^2}+\frac{df}{d\rho}+\lo\beta-\frac{\rho}{4}-\frac{k_\lambda^2}{4\rho}\ro
    f=0,
\end{equation}
where
\begin{equation}\label{4.3}
    \beta=\frac{\al E}{\sqrt{m^2-E^2}},\ \
    k_\lambda^2=(2j+1)^2+4\nu_\lambda-4\alpha^2.
\ee
    Notice that equation (\ref{4.3}) (but with another meaning of parameters $k_\lambda$)
    appears in the non-relativistic Hydrogen system \cite{Fock} and can be easily integrated.
    For the bound states, i.e., for $m^2>E^2$, its
    solutions can be expressed via degenerated hypergeometric
    function ${\cal F}(\tilde n,d,\rho)$ as
    \begin{equation}\label{4.4}
    f=C\rho^\frac{k_\lambda}{2}\exp\lo-\frac{\rho}{2}\ro{\cal
    F}\left(\frac{k_\lambda+1}{2}-\beta,k_\lambda+1,\rho\right),
\end{equation}
where $C$ is an integration constant.

Solutions (\ref{4.4}) are bounded at infinity provided the
argument $\tilde n=\frac{k_\lambda+1}{2}-\beta$ is a non-positive
integer, i.e., $\tilde n=-n'=0,-1,-2,\cdots$. Then from
(\ref{4.3}) we obtain the possible values of energy for bound
states: \be\label{4.5} E=m\lo 1+\frac{\al^2}{\lo \lo
n'+1/2+k_\lambda\ro^2-\al^2\ro^{\frac12}}\ro^{-\frac12}.
\end{equation}
Here $k_\lambda$ are parameters defined in expression (\ref{4.3}),
where $\nu_\lambda$ takes the values which coincide with the roots
of the characteristic equation for matrix $M$:
\begin{equation}\label{4.6}
    \det(M-\nu_\lambda I)=0,
\end{equation}
where $I$ is  the unit matrix of the appropriate dimension
$D=2s+1$ for $s\leq j$ or $D=2j+1$ for $j\leq s$.

Thus we present the exact values of energy levels for the Coulomb
system for the orbital particle having arbitrary half-integer
spin. However, formulae (\ref{4.5}), (\ref{4.3}) include parameter
$\nu_\lambda$ defined with the help of the algebraic equation
(\ref{4.6}) of order $D$ which can be solved in radicals for
$j\leq 3/2$ or $s=3/2$. For other values of $s$ and $j$ the
formula (\ref{4.6}) defines an algebraic equation whose order is
larger than 4, which does not have exact analytic solutions.  The
related possible values of $\nu_\lambda$ should  be calculated
numerically.

In the following section we find analytic expressions for
approximate solutions of (\ref{4.6}) and expand energy levels
(\ref{4.5}) in th power series of $(g\al)^2$.

\section{General discussion of energy spectrum}

First we note that if $g=2,$ and $ s=1/2$ then equation (\ref{4.5})
reduces to the exact Sommerfeld formula for energy levels of the
Dirac particle in the Coulomb field. Indeed, using (\ref{A2}) and
solving (\ref{4.6}) for $s=1/2,\ g=2$ we obtain
\[\nu_\lambda=\frac14+2\lambda\sqrt{\lo j+\frac12\ro^2-\al^2}, \ \
\ \ \lambda=\pm\frac12\] so that the related formula (\ref{4.5})
takes the following form \[E=m\lo1+\frac{\al^2}{\lo n'+\sqrt{\lo
j+\frac12\ro^2-\al^2}\ro^2}\ro^{-1/2}\] i.e., coincides with the
Sommerfeld formula. This seems to be rather curious since the
Dirac equation appears as a very particular case of equation
(\ref{2.3}) in which wave function
$\psi^{[\mu_1\nu_1]\cdots[\mu_n\nu_n]}$ has zero number of pairs
of anticommuting indices $[\mu_i\nu_i]$.  The second order
equation (\ref{3.3}) with $g=2$ and $\bf S$ being matrices of spin
1/2 , appears in the Dirac theory too if we use potential
(\ref{3.1}) and express the "small" components of the wave
function in terms of "large" ones (this procedure is equivalent to
our substitution (\ref{2.5})). Thus the spectrum of energies for
higher spin fermions (\ref{4.5}) includes the spectrum of the
Dirac particle as a particular case.

In order to analyze spectrum of energies of the orbital fermion
with arbitrary spin we shall look for approximate solutions of
equation (\ref{4.6}).  Considering $\al$ as a small parameter,
using explicit expressions (\ref{3.9}) and (\ref{A2}) for matrix
$M$ and applying the standard perturbation technique, we obtain
analytic expressions for approximate solutions of this equation
for arbitrary spin and total angular momenta:
\begin{equation}\label{5.1}\ba{l}
    \nu_\lambda=\lambda^2-(2j+1)\lambda\\\\\ds+\frac{(\al
    g)^2}{8}\lo\frac{(\lambda
    +s)(2j-\lbd-s+1)(s-\lbd+1)(2j+s-\lbd+2)}
    {(2j-2\lbd+1)(2j-2\lbd+3)(j-\lbd+1)}\right.\\ \ \\\ds\left.-
    \frac{(s+\lbd+1)(s-\lbd)(2j-s-\lbd)(2j+s-\lbd+1)}{(j-\lbd)(2j-2\lbd+1)(2j-2\lbd-1)}\ro+O(\al^4).\ea
\end{equation}

Starting with (\ref{4.5}) and using (\ref{5.1}) we find
approximate expressions for energy levels up to the terms of order
$\al^4$:
\begin{equation}\label{5.2}\ba{l}\ds
    \frac Em=
    1-\frac{\al^2}{2n^2}+\frac{3\al^4}{8n^4}-\frac{\al^4}{n^3(2l+1)}\\ \ \\\ds+\frac{g^2\al^4}{8n^3(2l+1)^2}\lo
    \frac{(j-l+s)(j+l-s+1)(l-j+s+1)(j+l+s+2)}{(l+1)(2l+3)}\right.\\ \ \\\ds\left.
    -\frac{(j-l+s+1)(j+l-s)(l-j+s)(j+l+s+1)(1-\delta_{l0})}{l(2l-1)}\ro,\\ \
    \\n=n'+l+1=1,2,\cdots,\
    l=0,1,\cdots,n-1.\ea\ee

Formula (\ref{5.2}) defines the fine structure of the energy
spectrum for particle with arbitrary spin. We see that energy
levels are labelled by three quantum numbers, i.e., by $n,l$ and
$j$. Any level of order $\al^2$ corresponding to the main quantum
number $n$ is split to $N_n$ sublevels which have the order
$\al^4$ and describe the fine structure of the spectrum. The value
of $N_n$ coincides with the number of possible pairs $(l,j)$,
i.e.,
\[N_n=\left\{ \ba{ll}n^2&\texttt{for}\ n\leq s,\\\lo s+\frac12\ro\lo
2n-s-\frac12\ro&\texttt{for}\ n>s.\ea\right.\]

We recall that for the Dirac particle the number of split
sublevels is equal to $n$ and any level with $j \neq n-1/2$ is
double degenerated. This again is in accordance with our formula
(\ref{5.2}) if we put $s=1/2$ and $g=2$ and find the related
approximate $\nu_\lambda$ from (\ref{5.1}).

 The physical interpretation of spectrum
 (\ref{5.2})
is obtained by using the results of paper \cite{NN} where the
Foldy-Wouthuysen reduction of equations (\ref{4.3}) was carried
out. In addition to the rest energy term proportional to $m$ the
formula (\ref{5.2}) includes the non-relativistic Balmer term
$-\frac{\al^2}{2n^2}$ and the term
$\al^4\lo\frac{3}{8n^4}-\frac1{n^3(2l+1)}\ro$ which represents the
relativistic correction to the kinetic energy. The last terms in
(\ref{5.2}) which are proportional to $g^2\al^4$, represent the
contribution caused by the spin-orbital, Darwin  and quadrupole
interactions.

\section{Energy levels for $s=3/2$}

In the previous section we have defined the energy levels for a
charged particle with arbitrary half-integer spin interacting with
the Coulomb potential and briefly described the related wave
functions. Here we consider in full detail the case $s=3/2$.

First we notice that in the case $s=3/2$ it is possible to solve
the corresponding algebraic equation (\ref{4.6}) in radicals and
to find the associated exact formulae (\ref{4.5}). We shall not
present these rather cumbersome formulae here and restrict
ourselves to analysis of approximate solutions (\ref{5.2}).

The energy levels (\ref{5.2}) for $s=3/2$ are reduced to the
following form

\begin{equation}\label{6.1}\ba{l}\ds
     E=
    m\lo 1-\frac{\al^2}{2n^2}+\frac{3\al^4}{8n^4}+\frac{\al^4}{n^3}\Delta_{j\ l}\ro.\ea\ee
Here $j$ is the total angular moment quantum number which can take
the following values
\begin{equation}\label{6.2}
    j=\left\{\ba{ll} \frac32,&\text{if} \ l=0,\\\frac52,\ \frac32,\
    \frac12,&\text{if}\ l=1,\\l+\frac32,\ l+\frac12,\ l-\frac12,\
    l-\frac32,&\text{if} \ l>1\ea\right.\ee
and $\Delta_{j\ l}$ are corrections to energy levels defined by
$j$ and $l$:
    \be\label{an2}\ba{l}\ds\Delta_{j\ l }=\frac{g^2}{8(2l+1)^2}\lo\frac{\left(j-l+\frac32\right)
    (j+l-\frac12)(l-j+\frac52)(j+l+\frac72)}{(l+1)(2l+3)}\right.\\\\\ds \left.
    -\frac{(j-l+\frac52)(j+l-\frac32)(l-j+\frac32)(j+l+\frac52)(1-\delta_{l0})}{l(2l-1)}\ro-\frac1{2l+1}.\ea\ee
    Thus in accordance with (\ref{6.2}), (\ref{an2}) we have
\begin{equation}\label{6.3}\ba{l}

    \ds\Delta_{l+ 3/2\ l}=\frac{3g^2(2l+5)}{8(l+1)(2l+1)(2l+3)}-\frac1{2l+1}\ \ \ \texttt{for any}
    \ l,\\
    \\
     \ds\Delta_{l+ 1/2\ l}=\frac{g^2(2l^2-9l-27)}{8l(l+1)(2l+1)(2l+3)}-\frac1{2l+1}\
    \ \  \texttt{for}\ l>0,\\
     \\
   \ds  \ds\Delta_{l- 1/2\ l}=\frac{g^2(16-2l^2-13l)}{8l(l+1)(2l+1)(2l-1)}-\frac1{2l+1}\
  \ \ \texttt{for}\ l>0,\\
\\
\ds\Delta_{l-3/2\
l}=\frac{3g^2(3-2l)}{8l(2l+1)(2l-1)}-\frac1{2l+1}\ \ \
\texttt{for}\ l>1.\ea
\end{equation}

Formulae (\ref{6.1})-(\ref{6.3}) describe the energy spectrum of a
particle of spin 3/2 interacting with the field of a point charge
up to order $\alpha^4$. We see that if $l>1$ then any level
corresponding to fixed values of the quantum numbers $n$ and $l$
is split to four sublevels representing a fine structure of the
spectrum. The value of the fine splitting decreases as the quantum
numbers $n$ and $l$ increase.

We notice that unlike the case $s=1/2$ energy levels (\ref{6.1})
are in general non-degenerate. The fine structure of spectrum
(\ref{6.1})-(\ref{6.3}) however includes an arbitrary parameter
$g$ which can be interpreted as a gyromagnetic ratio of the system
\cite{NN}. This parameter can be fixed by means of experimental or
theoretical requirements and some degeneracy can appear for its
special values.

There are three "privileged" values of $g$ namely the generally
recognized value $g=2$ (which is predicted by string theory and is
in accordance with experimental data \cite{tel}), $g=1/s$ (which
naturally appears in relativistic models without anomalous
interaction \cite{FN1}) and $g=\sqrt{2/s}$. For $g\neq\sqrt{2/s}$
it is possible to show that the spectrum (\ref{6.1})-(\ref{6.3})
is degeneracy free in general and for $g=2$ or $g=1/s$ even
accidental degeneracy free. However for $g=\sqrt{2/s}$ this
spectrum is doubly degenerate. Moreover, as we shall see, the
origin of this degeneracy is similar to one observed in the
Dirac-Coulomb system (cf. \cite{landau}).

Setting in (\ref{6.3}) $g=\sqrt\frac2s$ we obtain for $s=3/2$:
    \be\label{6.4}\ba{l}\ds\Delta_{l+\frac32\
    l}=-\frac{4l(l+2)+1}{2(2l+1)(l+1)(2l+3)}\ \texttt{for any}\ l,\\\\
    \ds\Delta_{l+\frac12 \
    l}=-\frac{4l^2(3l+7)+27(l+1)}{6l(l+1)(2l+1)(2l+3)},\ \ l>0,\\\\
    \ds\Delta_{l-\frac12\
    l}=-\frac{4l^2(3l+2)+7l-16}{6l(2l-1)(2l+1)(l+1)},\ \ l>0,\\\\
    \ds\Delta_{l-\frac32\ l}=-\frac{4l^2-3}{2l(2l+1)(2l-1)},\ \ l>1.\ea\ee

Comparing $\Delta_{l+\frac32\ l}$ with $\Delta_{l-\frac32\ l}$ in
(\ref{6.4}) it is not difficult to observe a double degeneracy of
the related energy levels, due to the fact that
    \be\ds\label{6.5}\Delta_{l+\frac32\
    l}=\Delta_{l-\frac12\ l+1},\ \ l\neq0.\ee

Thus if the gyromagnetic ratio is equal to $\sqrt\frac2s$ then
like in the Dirac-Coulomb system  the energy levels corresponding
to maximal or minimal possible value of $j$ for given $l>0$ are
doubly degenerate.  However, in some ways, this analogy is rather
matter of convention since in the Dirac-Coulomb system all energy
values (except the ground one) are doubly degenerate. In our
system with spin 3/2 like in the Dirac-Coulomb one there are
degenerate states with $j=l\pm s$. But in addition for $s=3/2$
there are states with $j=l\pm(s-1)$ and with $l=0$ which are
singlets.

We note that analysis of energy spectrum (\ref{5.2}) presented in
this section admits a straightforward extension to the case of any
spin. In particular, our conclusion concerning degeneracy of the
spectrum for special value $g=\sqrt{\frac2s}$ of the gyromagnetic
ratio is valid for any $s$.

\section{Explicit solutions of the relativistic wave equation for $s=\frac32$}

Following the procedure outlined in Sections 4 and 5 it is not
difficult to find the explicit expressions for the wave function
of the particle with spin 3/2 in the Coulomb field.

We start with the basic equation (\ref{2.3}). In the case $s=3/2$
the related wave function is a tensor-bispinor $\psi^{[\mu\nu]}$
which has 24 components, and equation (\ref{2.3}) is reduced to
the following form
\begin{equation}\label{7.0}\ba{l}
    \left(\gamma_\lambda \pi^\lambda-m\right)\psi^{[\mu\nu]}
-\frac{1}{6}\left(\gamma^{\mu}\gamma^{\nu}
-\gamma^{\nu}\gamma^{\mu}
\right)\pi_\lambda\gamma_\s\psi^{[\lambda\s]}\\
 \\+\frac{4iek}{3m}\lo\frac14\gm_\al\gm_\s
F^{\al\s}\psi_+^{[\mu\nu]}+
{F_\al}^{\mu}\psi_+^{[\nu\al]}-{F_\al}^{\nu}\psi_+^{[\mu\al]}\ro=0,\ea
\end{equation}
where (and in the following)
\be\label{7.01}\psi_\pm^{[\mu\nu]}=\psi^{[\mu\nu]}\mp
\frac12\gm_5{\varepsilon^{\mu\nu}}_{\lambda\s}\psi^{[\lambda\s]}.\ee

Solving RWE (\ref{7.0}) for $\psi^{[\mu\nu]}_-$ we obtain
equations (\ref{2.4}) and (\ref{2.5}) for particle with spin 3/2:
\begin{equation}\label{7.10}\ba{l}
\lo\pi_\al\pi^\al-m^2-\frac{g}{2}S_{\al\s}F^{\al\s}\ro\psi_+^{[\mu\nu]}=0
,\ea\ee
 \be\label{7.11}  \psi_-^{[\mu\nu]}
=\frac{1}{m} \gamma_\lambda\pi^\lambda\psi_+^{[\mu\nu] }.
\end{equation}

 Functions $\psi^{[\mu\nu]}_+$ and
$\psi^{[\mu\nu]}_-$ both have 12 independent components and form
carrier spaces for representations $D(3/2,0)\oplus D(0,3/2)\oplus
D(1/2,0)\oplus D(0,1/2)$ and $D(1,1/2)\oplus D(1/2,1)$ of the
Lorentz group respectively.

In addition, $\psi^{[\mu\nu]}$ has to satisfy the condition
(\ref{2.1}):
\begin{equation}\label{7.00}
    \gamma_\mu\gamma_\nu\psi^{[\mu\nu]}
=0.
\end{equation}

In accordance with its definition, tensor $\psi^{[\mu\nu]}_-$
automatically  satisfies the condition
$\gm_{\mu}\gm_{\nu}\psi^{[\mu\nu]}_-=0$ (even when condition
(\ref{7.00}) for $\psi^{[\mu\nu]} $ is not imposed). Function
$\psi^{[\mu\nu]}_+$ satisfies condition
$\gm_{\mu}\gm_{\nu}\psi^{[\mu\nu]}_+=0$ as well
 provided equation (\ref{7.00}) is satisfied. Moreover, relation (\ref{7.00}) reduces
 the number of independent components of
$\psi^{[\mu\nu]}_+$ to 8 which form a carrier space of the
representation $D(3/2,0)\oplus D(0,3/2)$.

The action of the generators of the Lorentz group on
$\psi^{[\lambda\s]}_+$ can be derived from equation (\ref{2.6})
for $n=1: $\begin{equation}\label{7.000}\ba{l}
    S^{\mu\nu}\psi_+^{[\lambda\s]}
=\frac{i}{4}[\gm^\mu,\gm^\nu]\psi_+^{[\lambda\s]}+i\lo
g^{\mu\lambda}\psi_+^{[\nu\s]} -g^{\nu\lambda}\psi_+^{[\mu\s]}
-g^{\mu\s}\psi_+^{[\nu\lambda]}+g^{\nu\s}\psi_+^{[\mu\lambda]}\ro.\ea
\end{equation}
In fact the same action is valid for any tensor-bispinor, e.g.,
for $\psi^{[\lambda \s]}$ and $\psi^{[\lambda \s]}_-$. For
$\psi^{[\lambda \s]}_+$ the formula (\ref{7.000}) can be specified
taking into account equations (\ref{7.01}) and (\ref{7.00}). First
we note that in accordance with (\ref{7.01})
\be\label{7.02}\psi^{[0a]}_+=\frac12{\va^{0a}}_{bc}\gm_5\psi^{[bc]}_+\
\ (a,b,c =1,2,3)\ee so that we can restrict ourselves to spatial
components $\psi^{[ab]}_+$ of the tensor-spinor with $a, \
b=1,2,3$ (we denote superindices running from 1 to 3 by Latin
letters). Denoting \be\label{7.0000}
\psi^{[ab]}_+=\va^{abc}\eta^c\ee and using the relations
(\ref{7.01}), (\ref{7.00}), (\ref{7.000}) and (\ref{7.02}) we
obtain the action of generators of the Lorentz group in the
following form:
    \be\label{7.03}\ba{l}S^{ab}\eta^c=\va^{abk}S^k\eta^c,\ \
    S^{0a}\eta^c=i\gm_5S^a\eta^c,\ea\ee
    where
    \be\label{7.04}S_a=S^{(\frac12)}_a+ S^{(1)}_a,\ \ \
    S^{(\frac12)}_a=\frac14\va_{abc}\gm_b\gm_c\ee
    and $S^{(1)}_a$ are spin one matrices whose elements are
    $\lo  S^{(1)}_a\ro_{bc}=i\va_{abc}$.

We note that like tensor-spinor $\psi^{[\mu\nu]}_+$, vector-spinor
$\eta^a\equiv\eta^a_\al$ has an implicit spinorial index $\al$
which we usually omit. Moreover, $\gm$-matrices act on spinorial
index of $\eta^a$ while matrices $S^{(1)}_a$ act on vector index
$a$ so that $S^{(1)}_a$ commute with $\gm_\mu$ by definition.
    Thus matrices $S_a$ in (\ref{7.04}) are sums of commuting spin 1/2 and spin 1
    matrices, and so that they can be reduced to the direct sums of spin 3/2
    and spin 1/2 matrices. Such reduction is easily handled my means of
     Wigner coefficients provided the corresponding third
    components of spin vectors are diagonal.

       If we choose the following realization
    of the
    Dirac matrices:
        \be\label{7.05}\gm_0=\lo\ba{rr}I&\hat 0\\\hat 0&-1\ea\ro, \ \
        \gm_a=\lo\ba{ll}\hat 0&-\s_a\\\s_a&\ \ \hat 0\ea\ro,\ \
        i\gm_5=\lo\ba{rr}I&\hat 0\\\hat 0&-I\ea\ro\ee
where $I$ and $\hat 0$ are the $2\times 2$ unit and zero matrices
respectively and $\s_a$ are the Pauli matrices, then both $i\gm_5$
and $S^{(\frac12)}_3$ are diagonal. Using the unitary
transformation $\eta^a\to W^{ab}\eta^b$ with the transformation
matrix
\be\label{7.06}W=\frac1{\sqrt2}\lo\ba{ccc}-i&0&i\\1&0&1\\0&i\sqrt2&0\ea\ro\ee
we diagonalize $S^{(1)}_3$ too since in this case
\[S^{(1)}_3\to \tilde S^{(1)}_3=WS^{(1)}_3W^\dag=\lo\ba{rrr}1&0&0\\0&0&0\\0&0&-1\ea\ro.\]

Thus we pass from (\ref{7.04}) to the following realization of
spin matrices
    \be\label{7.07}S_a\to \hat S_a=WS_aW^\dag=\frac12\s_a+\tilde
    S^{(1)}_a.\ee

Now we are ready to apply the Wigner coefficients
$C^{sm}_{{s_1}{m_1}{s_2}{m_2}}$ to reduce (\ref{7.07}) to a direct
sum of spin 3/2 and spin 1/2 matrices. In our case $s_1=1,
s_2=1/2, s=3/2,1/2$ and the following notation is used: $m=k, \
m_1=c, \ m_2= \al$. Then the basis $\Psi^{sk}$ which corresponds
to the completely reduced representation of spin matrices is
connected with $\eta^a_\al$ by means of the formula:
\be\label{7.08}\Psi^{sk}=C^{sk}_{1c\frac12\al}W_{ca}\eta^a_\al.\ee

Condition (\ref{2.1}) suppresses the states with $s=1/2$ so that
$\Psi^{\frac12k}=0$. Then, denoting $\Psi^{\frac32k}=\Psi^k$ and
using (\ref{7.08}) we obtain:
\be\label{7.09}\ba{l}\Psi^k=C^{\frac32k}_{1c\frac12\al}W_{ca}\eta^a_\al.\ea\ee

Thus relations (\ref{7.02}), (\ref{7.0000}) and (\ref{7.09})
explicitly show how functions $\psi^{[\mu\nu]}_+$ (which satisfy
equations (\ref{7.10})) are connected with solutions $\Psi^k$ of
the reduced equation (\ref{2.62}). The inverse relations have the
form
    \be\label{7.12} \psi^{[ab]}_{\al+}=\va_{abc}\lo
    W^\dag\ro_{ck}C^{\frac32n}_{1k\frac12\al}\Psi^n, \ \psi^{[0c]}_{\al+}=\lo\gm_5\ro_{\al\s}\lo
    W^\dag\ro_{ck}C^{\frac32n}_{1k\frac12\s}\Psi^n.\ee

In accordance with expressions (\ref{n0})--(\ref{4.4}) solutions
of the reduced equations (\ref{2.62}) corresponding to energy
$E=E_{n' j\ \nu_\lambda}$ (\ref{4.5}) can be written as
\be\label{7.1}\Psi^k_{n' j\ \nu_\lambda}=\exp(-iE_{n' j\
\nu_\lambda}x_0)\om_\lambda\tilde\xi_{\lambda}\Omega^\frac32_{j\
j-\lambda\ m},\ee
 where  $\om_\lambda$ are the eigenvectors of matrix $M$ whose elements
$M_{\lambda\lambda'}$ are given by formula (\ref{3.9}) (see
(\ref{A5})), and: \be\label{7.2}\ba{l}
\tilde\xi_\lambda=C_\lbd(m^2-E^2_{n' j\
\nu_\lambda})^\frac{k_\lambda+1}4r^\frac{k_\lambda-1}2e^{(m^2-E^2_{n'
j \ \nu_\lambda})r}{\cal F}\lo-n', k_\lambda+1, 2\sqrt{
m^2-E^2_{n' j\ \nu_\lambda}}r\ro.\ea\ee

Substituting $\Psi^k=\Psi^k_{n' j\ \nu_\lambda}$ from (\ref{7.1})
into (\ref{7.12}) we obtain the explicit expressions for
(non-normalized) solutions of the equation (\ref{7.0}).

To complete presentation of solutions of the reduced equation
(\ref{2.62}) for spin $s=3/2$ we present the explicit expressions
of spherical spinors $\Omega^\frac32_{j\ j-\lambda\ m}$ in the
Appendix.

\section{Discussion}

In the paper the explicit solution of the relativistic quantum
mechanical Kepler problem for orbital particle with arbitrary
half-integer spin are presented. As a mathematical model of such a
system we use the tensor-spinor relativistic wave equations (RWE)
proposed in paper \cite{NN}.

Tensor-spinor formulation of RWE makes it possible to overcome the
main fundamental difficulties which appears in relativistic theory
of particles with higher spins like causality violation and
complex energies predicted for particles interacting with the
constant magnetic field \cite{NN}. In addition, this formulation
is rather straightforward and simple which makes it possible to
extend the known exact solutions for some special quantum
mechanical problems described by the Dirac equation \cite{bagr} to
the case of an arbitrary half-integer spin. The problem of
interaction of arbitrary spin fermions with constant and
homogeneous magnetic field was solved in paper \cite{NikGal}.

 In the
present paper we solve a much more complicated problem:
interaction of arbitrary spin fermion with the field of a point
charge. In addition, tensor-spinor RWE admit exact solutions for
the cases of interaction with the plane wave field, with constant
crossed electric and magnetic field and some others. We plan to
present solutions of the related problems in the future.

In addition to the general treatment for an arbitrary spin we
restrict in full details to case $s=3/2$. However the results
given in Sections 6 and 7 admit straightforward  generalization to
the case of $s$ arbitrary. In particular, in analogy with
(\ref{7.02})-(\ref{7.12}) the tensor-spinor $\psi_+^{[\mu_1\nu_1]
[\mu_2\nu_2]\cdots[\mu_n\nu_n]}$ with arbitrary $n$ in view of
equation (\ref{2.1}) is effectively reduced to symmetric tensor
$\eta^{\mu_1\mu_2\cdots\mu_n}$ with $n$ indices running from 1 to
3, and then to a $2(2s+1)$-component function $\psi$ satisfying
equation (\ref{3.3}).

We obtain the generalized Sommerfeld formula (\ref{4.5}) for
energy levels of particle with arbitrary half-integer spin
interacting with the Coulomb potential. In contrast with the
formula generated by the Dirac equation our expression (\ref{4.5})
includes parameter $g$ whose value is not fixed {\it a priori}. In
accordance with the analysis present in \cite{NN} this parameter
is associated with the gyromagnetic ratio of a particle described
by equations (\ref{2.1}), (\ref{2.3}).

Analyzing formula (\ref{5.2}) for energy levels we conclude that
in addition to the most popular values $g=1/s$ and $g=2$ there
exist one more intriguing value, namely $g=\sqrt{2/s}$ which
corresponds to a specific degeneracy of the related energy
spectrum.
 We notice that in the case $s=1/2$ all mentioned privileged values
 of the
gyromagnetic ratio coincide while for $s>1/2$ the relation 
\[\frac1s<\sqrt\frac2s<2\] is valid. In other words the degeneracy related value
of $g$ lies between the recognized values $1/s$ and 2. 

\vspace{5mm}

One of us (AGN) is indebted to the Ministery of Education, Youth and Sports of the
Czech republic for the support throw 
the grant number 1P2004LA211.


\appendix
\section{Appendix}

Here we present some technical data used in the main text.

Spherical spinors are eigenvectors of the commuting operators of
squared total angular momentum ${\bf J}^2$, squared orbital
momentum ${\bf L}^2$, squared spin ${\bf S}^2$ and the third
components $J_3$ of total angular momentum, so that
\[\ba{l}{\bf J}^2\Omega^s_{j\ l\ m}=j(j+1)\Omega^s_{j\ l\ m},\ \
{\bf L}^2\Omega^s_{j\ l\ m}=l(l+1)\Omega^s_{j\ l\ m},\\{\bf
S}^2\Omega^s_{j\ l\ m}=s(s+1)\Omega^s_{j\ l\ m},\ \
J_3\Omega^s_{j\ l\ m}=m\Omega^s_{j\ l\ m}.\ea\]

Spherical spinors can be expressed via spherical functions
$Y_{lm}$:
\[\lo\Omega^s_{j\ l\ m}\ro_\mu=C^{jm}_{l\ m-\mu\ s\ \mu}Y_{l\ m-\mu}\]
where $C^{jm}_{l\ m-\mu\ s\ \mu}$ are Wigner coefficients,
\[\ds\ba{l}Y_{lm}\lo\frac{\bf
r}{r}\ro=\frac1{2\pi}\exp(im\vp)(-1)^\frac{m+|m|}{2}\lo\frac{(l+1/2)
(l-|m|)!}{(l+|m|)!}\ro^{1/2}P_l^{|m|}(\cos(\theta),\\
\ds
P_l^{|m|}(\cos\theta)=\frac{1}{l!2^l}\sin^{|m|}\theta\frac{d^{|m|+1}}{(d\cos\theta)^{|m|}}
(\cos^2\theta-1)^l,\ea\] $\vp$ and $\theta$ are polar and
azimuthal angles of $\bf r$.

The result of action of matrix ${\bf S}\cdot{\bf r}$ to spherical
spinors is given by relation (\ref{3.5}) where \cite{MH},
\cite{FN1}
\be\label{A2}\ba{l}K^{sj}_{\lambda\lambda'}=-1/2\lo\delta_{\lambda'\lambda+1}\
a^{sj}_{s+ \lambda}+\delta_{\lambda'\lambda-1}\
a^{sj}_{s+\lambda+1}\ro,\\\\
\ds a^{sj}_\mu=\lo\frac{\mu(2j+1-\mu)(2s+1-\mu)(2j+2s-\mu+2)}
{(2s+2j-2\mu+1)(2s+2j-2\mu+3)}\ro^{1/2}.\ea\ee

In particular, for $s=3/2$, the spherical spinors can be
represented as the following four-component rows:
\[\ba{l}\Omega^\frac32_{j\
j-\frac32\
m}=\frac1{2\sqrt{j(j-1)(2j-1)}}\left(\ba{c}\sqrt{(j+m)(j+m-1)(j+m-2)}Y_{j-\frac32\
m-\frac32}\\\sqrt{3(j^2-m^2)(j+m-1)}Y_{j-\frac32\
m-\frac12}\\\sqrt{3(j^2-m^2)(j-m-1)}Y_{j-\frac32\
m+\frac12}\\\sqrt{(j-m)(j-m-1)(j-m-2)}Y_{j-\frac32\
m+\frac32}\ea\ro,\ea\]
\[\ba{l} \Omega^\frac32_{j\ j-\frac12\
m}=\frac1{2\sqrt{j(j+1)(2j-1)}}\left(\ba{c}-\sqrt{3(j+m)(j+m-1)(j-m+1)}Y_{j-\frac12\
m-\frac32}\\-(j+1-3m)\sqrt{j+m}Y_{j-\frac12\
m-\frac12}\\(j+1+3m)\sqrt{j-m}Y_{j-\frac12\
m+\frac12}\\\sqrt{3(j-m)(j-m-1)(j+m-1)}Y_{j-\frac12\
m+\frac32}\ea\ro,\ea \]
\[\ba{l} \Omega^\frac32_{j\ j+\frac12\
m}=\frac1{2\sqrt{j(j+1)(2j+3)}}\left(\ba{c}\sqrt{3(j+m)(j-m+2)(j-m+1)}Y_{j+\frac12\
m-\frac32}\\-(j+3m)\sqrt{j-m+1}Y_{j+\frac12\
m-\frac12}\\-(j-3m)\sqrt{j+m+1}Y_{j+\frac12\
m+\frac12}\\\sqrt{3(j-m)(j+m+1)(j+m+2)}Y_{j+\frac12\
m+\frac32}\ea\ro,\ea\]\[\ba{l} \Omega^\frac32_{j\ j+\frac32\
m}=\\=\frac1{2\sqrt{(j+1)(j+2)(2j+3)}}\left(\ba{c}-\sqrt{(j-m+1)(j-m+2)(j-m+3)}Y_{j+\frac32\
m-\frac32}\\\sqrt{3(j+m+1)(j-m+1)(j-m+2)}Y_{j+\frac32\
m-\frac12}\\-\sqrt{3(j-m+1)(j+m+1)(j+m+2)}Y_{j+\frac32\
m+\frac12}\\\sqrt{(j+m+1)(j+m+2)(j+m+3)}Y_{j+\frac32\
m+\frac32}\ea\ro \ea\] and the coefficients (\ref{A2}) are reduced
to the following form:
    \be\label{A10} a^{\frac32 j}_{1}=\sqrt{\frac{3(j+1)}{j}},\ \
    a^{\frac32 j}_{2}=\sqrt{\frac{(2j-1)(2j+3)}{j(j+1)}},\ \
    a^{\frac32 j}_{3}=\sqrt{\frac{3j}{j+1}}.\ee

   The related matrix $M$ whose
elements $M_{\lambda\lambda'}$ are given by formula (\ref{3.9})
takes the form:
\be\label{A4}M=\lo\ba{llll}\frac{15}{4}+3j&\frac{ig\al}{2}\sqrt\frac{3j}{j+1}&0&0\\
\frac{ig\al}{2}\sqrt\frac{3j}{j+1}&\frac34+j&\frac{ig\al}{2}\sqrt{\frac{(2j-1)(2j+3)}{j(j+1)}}&0\\
0&\frac{ig\al}{2}\sqrt{\frac{(2j-1)(2j+3)}{j(j+1)}}&-\frac14
-j&\frac{ig\al}{2}\sqrt{\frac{3(j+1)}{j}}\\
0&0&\frac{ig\al}{2}\sqrt{\frac{3(j+1)}{j}}&\frac34-3j \ea\ro.\ee

Eigenvalues $\nu$ of matrix (\ref{A4}) coincide with the roots of
the characteristic equation for matrix (\ref{A4}) which is an
algebraic equation of order four:
\be\label{A11}\ba{l}2\nu^4-10\nu^3+\lo\frac{39}{4}+5(g\al)^2-10j(j+1)\ro\nu^2\\+\lo
18j(j+1)-\frac98-\frac{57}{2}(g\al)^2\ro\nu+18j^2(j+1)^2\\-\frac94
j(j+1)-\frac52\lo\frac34\ro^4-\lo 9j(j+1)-48.75\ro(g\al)^2=0.\ea
\ee

The related eigenvectors $\om_\lambda$ corresponding to
eigenvalues $\nu_\lambda,\ \lambda=1,2,3,4$ can be observed in the
following form
    \be\label{A5}\om_\lambda=\lo\ba{c}\frac{(g\al)^2\sqrt{3j(2j-1)(2j+3)}}{j+1}
    (4\nu_\lambda+12j-3)
        \\
\frac{i(g\al)^2\sqrt{(2j-1)(2j+3)}}{2\sqrt{j(j+1)}}({15}+12j-4
\nu_\lambda)(3-12j-4\nu_\lambda)
    \\\lo(15+12j-4
\nu_\lambda)(3+4j-4\nu_\lambda)+
\frac{12(g\al)^2(j+1)}{j}\ro(4\nu_\lambda+12j-3)
\\\frac{ig\al\sqrt{3(j+1)}}{2\sqrt{j}}\lo(15+12j-4
\nu_\lambda)(3+4j-4\nu_\lambda)+
\frac{12(g\al)^2(j+1)}{j}\ro\ea\ro\ee

Relation (\ref{A5}) is used in formula (\ref{7.1}).

Solving algebraic equation (\ref{A11}) it is possible to find
exact values of $\nu_\lambda$. We shall not present the related
cumbersome formulae here whose expansion in power series of
$(g\al)^2$ is given in formula (\ref{5.1}).

\end{document}